# Experimental Demonstration of 4,294,967,296-QAM Based Y-00 Quantum Stream Cipher Carrying 160-Gb/s 16-QAM Signals

**XI CHEN,**[1,*] **KEN TANIZAWA,**[2] **PETER WINZER,**[3] **PO DONG,**[3] **JUNHO CHO,**[1] **FUMIO FUTAMI,**[2] **KENTARO KATO,**[2] **ARGISHTI MELIKYAN,**[1] **AND KW KIM**[1]

[1]*Nokia Bell Labs, Holmdel, New Jersey, NJ 07733, United States*
[2]*Quantum ICT Research Institute, Tamagawa University, 6-1-1 Tamagawa Gakuen, Machida, Tokyo, 194-8610, Japan*
[3]*Was with Nokia Bell Labs when involved in this project*
*\* xi.v.chen@nokia-bell-labs.com*

**Abstract:** We demonstrate a 4,294,967,296-ary quadrature amplitude modulation (QAM) based Y-00 quantum stream cipher system carrying 160-Gb/s 16-QAM signal transmitted over 320-km SSMF. The ultra-dense QAM cipher template is realized by an integrated two-segment silicon photonics I/Q modulator.

## 1. Introduction

Fiber-optic transmission systems are susceptible to eavesdropping and require encryption for sensitive data. Fiber-optic cables are typically readily accessible and can be tapped using, e.g., evanescent coupling at an intentionally introduced fiber bend [1]. So far, the only known way to achieve *perfect security* which is fundamentally immune to all kinds of attacks is the one-time-pad (OTP) based on a fundamentally secure key, as may be established using quantum key distribution (QKD) [2]. However, OTP encryption requires the key to have the same length as the data/plaintext, which limits the data rate to the key distribution rate. Instead of implementing a *perfectly secure* system, *practically secure* schemes are often used. For instance, the Advanced Encryption Standard (AES) is a practically secure scheme applied in application layer. With AES, security attacks (e.g. brute force attacks on the key) are possible but are computationally hard. Besides the higher layer encryption, physical-layer encryption [3-9] that combines *unavoidable* physical randomness (noise) with suitable algorithms can also allow *practical security*. For instance, the Y-00 quantum stream cipher [3,6-8] is one of the physical layer encryption schemes that uses inevitable quantum noise/shot noise at detection to mask the information. The scheme relies on an encryption algorithm to convert a lower-order quadrature amplitude modulation (QAM) original constellation to an ultra-dense higher-order QAM constellation, which together with fundamental shot noise in photodetection makes recovery of the information impossible without the key. A crucial design parameter of the Y-00 cipher scheme is the encrypted QAM size. The higher the order of the ultra-dense modulation is chosen that the original data is converted to, the more security the system can provide. So far, the largest demonstrated encrypted QAM size is 1,048,576 ($=2^{20}$) QAM with low-speed 10-bit DACs at a symbol rate of 5 GBaud and line rate of 70 Gb/s [9].

In this paper, we design an integrated two-segment silicon photonics (SiPh) I/Q modulator which allows cascaded modulation via two sets of DACs and drivers. This dramatically increases the modulation order and greatly enhances the security of the Y-00 scheme. We show that we can generate a 4,294,967,296 ($=2^{32}$) QAM based Y-00 ciphertext that carries 22-GBaud 16-QAM information at 160-Gb/s physical-layer secure net rate with 320-km standard single mode fiber (SSMF) transmission.

## 2. Y-00 Stream Cipher and Device Design

The Y-00 quantum stream cipher achieves secure data transmission via two steps: i) at the transmitter, an encryption algorithm with a short seed key (e.g. 256 bits) is used to convert a low-order QAM signal to an ultra-dense QAM waveform; ii) upon photodetection, shot noise introduces detection uncertainties on the waveform. With a sufficient number of amplitude levels on each quadrature of the ultra-dense QAM constellation, the noise induced uncertainties can lead to a symbol error ratio (SER) closely approaching 1 [10]. Given the coding of data bits is cyclically shifted and there is sufficient randomization on the amplitudes in the Y-00 encryption algorithm, SER of 1 in such a system will result a bit error ratio (BER) closely approaching 0.5 [11], meaning that the eavesdropper (Eve) has no other choice except for pure guessing. Only the legitimate user (Bob) who has the key can decrypt the waveform to the original low-order QAM signal for error-free detection. We assume that the short seed key is safe and can be delivered via a separate, highly secure distribution system such as QKD. The conversion from low-order QAM (e.g. 16-QAM in this demonstration) to ultra-dense QAM (e.g. 4,294,967,296-QAM) is done as illustrated in Fig. 1 (a). The seed key is sent into two pseudorandom number generators (PRNGs) to generate two sets of practically unrepeated key streams (e.g. a 256-bit seed key generates a keystream with a period of $2^{256}$-1). The key stream from PRNG1 runs an exclusive OR (XOR) logic on the plaintext, and the one from PRNG2 is used for randomizing the amplitude levels via a basis combination algorithm. Details of the algorithm can be found in Ref. [7]. For each polarization, the original 4-bits 16-QAM plaintext is randomized into 32 bits and generate a 4,294,967,296 (= $2^{32}$) QAM ciphertext symbol.

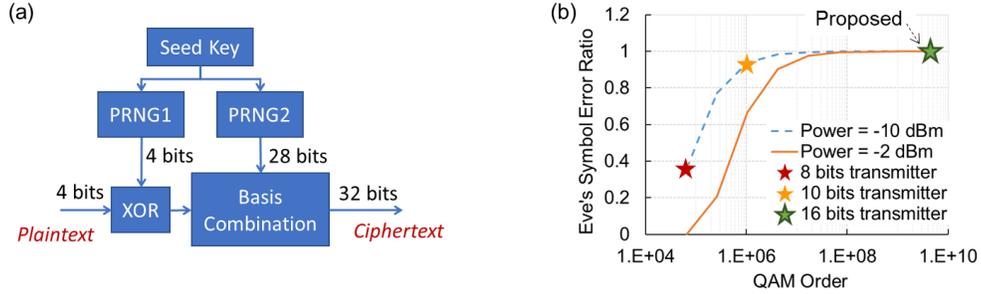

Fig. 1 (a) Block diagram for the Y-00 encryption algorithm; (b) Theoretical calculation on Eve's SER as a function of the QAM size.

To understand how large the QAM size needs to be in order to be masked by shot noise, a theoretical calculation is done following Ref. [12], with results shown in Fig. 1 (b). We examine Eve's SER at two critical power levels, i) -10 dBm which is a typical power level after an optical modulator; and ii) -2 dBm which is the optimal launch power for our 160 Gb/s signal at the begging of each fiber span. It can be seen that with only shot noise, an Eve's SER can be made to only 0.36 at the highest by an 8-bit transmitter that can generate $2^8 \times 2^8$ QAM at most (the red star in Fig. 1 (b)). A 10-bit transmitter (the orange star in Fig. 1 (b)) would help increasing the SER, which is however still limited to 0.93. With our two-segment SiPh I/Q modulator, the nominal number of bits can be doubled. In other words, two 8-bit DACs can be used for each quadrature to produce a 16-bit transmitter which allows a cipher template of 4,294,967,296 (= $2^{(8+8)} \times 2^{(8+8)}$) QAM.

A schematic diagram of the two-segment modulator is shown in Fig. 2 (a). As seen, the I/Q modulator consists of two Mach-Zehnder modulators (MZMs). Each MZM is formed by two waveguide phase shifters with embedded reverse-biased PN junctions in a push-pull configuration. The electrode for the phase shifter consists of two segments driven by two independent DACs. Each segment is 3 mm long and has a $V_\pi$ of ~6V. The inset to Fig. 2 (a) shows an example of modulating QPSK on each segment to generate a 16-QAM signal. A photo of the modulator chip is shown in Fig. 2 (b).

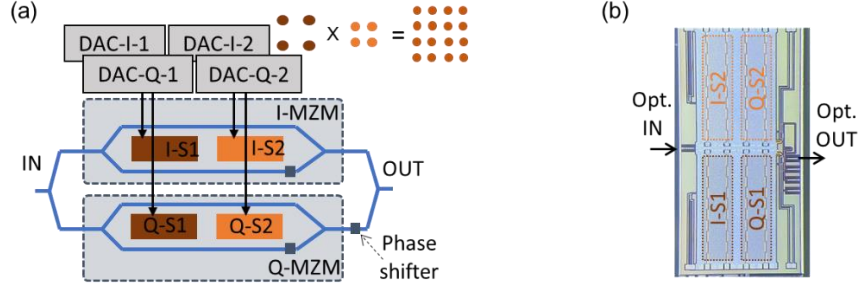

Fig. 2 (a) Schematic diagram of the two-segment SiPh I/Q modulator; (b) a photo of the modulator chip.

## 3. Experimental Setup

With the two-segment I/Q modulator, we demonstrate the transmission of a 22-GBaud 16-QAM signal encrypted into a 4,294,967,296-QAM Y-00 quantum stream ciphertext. The experimental setup is shown in Fig. 3. The transmitter consists of a 1-kHz linewidth laser operating at 1550.1 nm [13] and the two-segment SiPh I/Q modulator. The modulator uses a laser input power of 17.5 dBm and the transmitter generates an output power of -16.5 dBm. The high loss is due to the packaging and the relatively high $V_\pi$ of this experimental device. The DACs have a nominal resolution of 8 bits, and operate at 88 GSa/s. The electrical driver amplifiers have a gain of 23 dB. Polarization-division multiplexing (PDM) is realized by a fiber-delay based PDM emulator with 10.9 meters of decorrelation delay (i.e., 54.5 ns or 1199 symbols). The transmission fiber is 4 spans of 80-km dispersion-uncompensated SSMF. The signal launch power is -2 dBm. The inset of Fig. 3 shows the signal spectrum at the transmitter (blue) and the receiver (orange). The received signal is detected by an integrated coherent receiver (ICR) [14] with input optical power of –10 dBm. The optical signal-to-noise ratio (OSNR) before and after fiber transmission is 36 dB and 29.3 dB. The free-running local oscillator has a 1-kHz linewidth and is kept to within a frequency offset of ~300 MHz relative to the transmit laser. The ICR has integrated transimpedance amplifiers (TIAs) and generates electrical signals with peak-to-peak voltage of ~320 mV. The electrical signal is sampled by a real-time scope operating at 80 GSa/s with 8 nominal bits and electrical bandwidth of 32 GHz.

The digital signal processing (DSP) procedures are shown in Fig. 4 (a) and (b). At the transmitter, every 4 information bits (one 16-QAM symbol) generates a 32-bits ($2^{32}$-QAM) constellation point. No Nyquist pulse shaping is used, as it increases the required number of bits from the DACs. We aim to utilize all the 16-bit resolution for the ultra-dense QAM encryption. For each quadrature, the 16 bits [$b_{16}$, $b_{15}$, …, $b_2$, $b_1$] for the in-phase part of the $2^{32}$-QAM waveform are split into two 8-bit groups, the even-indexed bits [$b_{16}$, $b_{14}$…, $b_2$] and the odd-indexed bits [$b_{15}$, $b_{13}$, …, $b_1$] for two DACs. This results in 6-dB higher RF driving power on the 1st segment than on the 2nd segment. The RF power on the two segments are carefully adjusted by RF attenuators and gain-tunable drivers. At the receiver, coherent DSP such as dispersion compensation and multiple-input multiple-output (MIMO) adaptive filtering are done. The MIMO filter has 121 taps, and uses least mean square (LMS) algorithm with a digital phase locked (digital PLL) inside. The first 5000 symbols are used for pre-convergence, and then 3% pilot symbols are used for assisting the channel and phase estimation. After the MIMO filtering, the Y-00 decryption recovers the 16-QAM and binary bits in a symbol-by-symbol amplitude subtraction fashion [7] as illustrated in Fig. 4 (c). A BER is calculated with 9.46 million received bits of the 16-QAM user information.

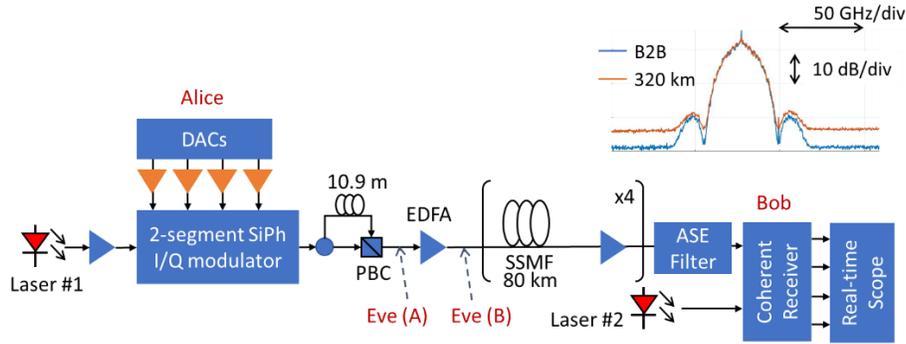

Fig. 3. Experimental setup for 22-GBaud 16-QAM signal with 4,294,967,296-QAM Y-00 quantum stream cipher encryption.

Note that our DACs have a limited memory of $2^{18}$ samples which is significantly less than the number of points in a $2^{32}$-QAM constellation. The $2^{32}$-QAM constellation therefore contains only a random subset of the $2^{32}$ possible constellation points within this experiment. We emphasize that the encryption security of this work is ensured by the massive number and density of all *possible* points in the cipher constellation template, and this is independent of what subset of the points is actually selected by encryption in our measured time window. In fact, our encryption scheme generates ultra-dense constellation points uniformly at random, and our measurement confirms that loading different random subsets of $2^{32}$ constellation points affect neither the BER for Bob's 16-QAM signal nor the SER for Eve's $2^{32}$-QAM waveform. In a real system where the bit stream has no cycle, measuring the whole $2^{32}$-QAM constellation can be done by increasing the observation time.

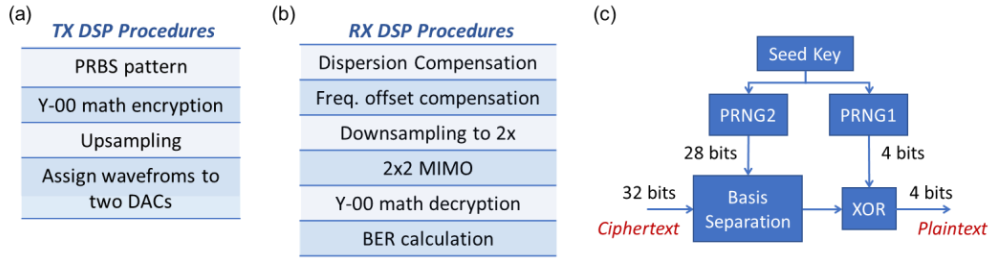

Fig. 4. Digital signal processing procedures for (a) transmitter and (b) receiver; (c) block diagram for the Y-00 decryption algorithm.

## 4. Results and Discussion

We keep the cipher template at $2^{32}$ QAM and measure the BER of the 22-GBaud 16-QAM signal as a function of ONSR at optical back-to-back setting. The results are shown in Fig. 5. A regular 22-GBaud 16-QAM signal without encryption is also measured for comparison. Assuming 7% hard-decision forward error correction (HD-FEC) with a BER threshold of $3.9 \times 10^{-3}$, the required OSNRs are 23 dB for the regular 16-QAM signal and 24 dB for the encrypted 16-QAM. The resulting net data rate is (22 GBaud $\times$ 16 QAM $\times$ 2 pol. /(1 + 7% FEC overhead + 3% pilot)) = 160 Gb/s. Note that the Y-00 encryption scheme can also work with soft-decision FEC (SD-FEC), if FEC decoding is incorporated in the basis separation function. The constellations before and after the decryption at an OSNR of 36 dB are shown in Fig. 5. The BER after 320-km SSMF transmission is $2.5 \times 10^{-3}$, shown as the red dot in Fig. 5.

In order to verify the effectiveness of noise masking with our increased QAM size, we measure the SER for Eve as a function of QAM size. We give Eve access to the two best places for tapping (c.f. Fig. 3): i) Point A, after data modulation and before the first EDFA, where the optical power may be weak but the amplified spontaneous emission (ASE) noise is the smallest; and ii) Point B, after the first EDFA where the optical power is strong but the signal is corrupted by ASE. The measured SER is shown in Fig. 6. We can observe that, although our system is impaired not only by shot noise but also by other noise such as thermal noise and analog-to-digital (ADC) quantization noise, no cipher template smaller than $2^{24}$ QAM can make Eve's SER higher than 99.98%. A $2^{24}$ QAM template requires 12 bits per quadrature, which exceed what commercial high-speed CMOS DACs can offer.

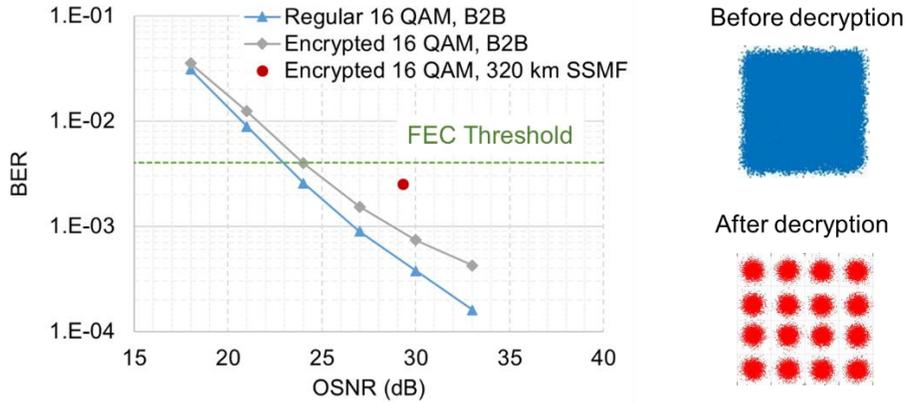

Fig. 5. Measured BER as a function of OSNR for the 22-GBaud 16-QAM signals with and without the encryption.

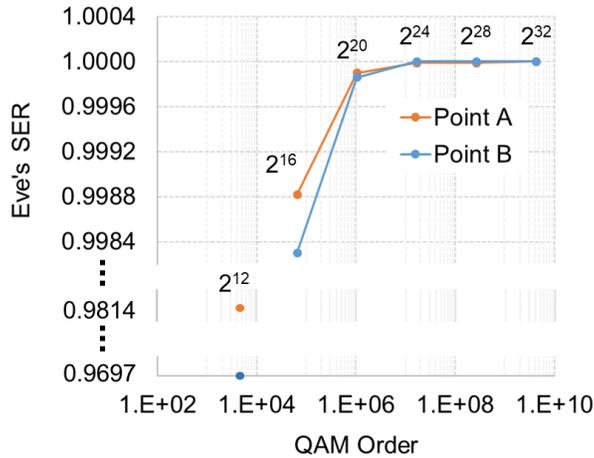

Fig. 6. Measured symbol error ratio as a function of QAM template size at eavesdropping point A & B.

## 5. Conclusions

We demonstrated a 4,294,967,296-QAM based Y-00 quantum stream cipher system carrying a 160-Gb/s net rate physical-layer secure signal. The ultra-dense QAM cipher signal generation is realized by our two-segment SiPh I/Q modulator driven by two sets of independent DACs with an aggregate nominal resolution of 16 bits per quadrature. The demonstrated encrypted

QAM template is $(2^{32}/2^{20} =)$ 4,096 times larger than the previous demonstration ($2^{20}$ QAM), and the symbol rate is (22 GBaud/5 GBaud =) 4.4 times higher than the previous demonstration [9].

## Disclosures

The authors declare no conflicts of interest.